\documentclass[conference]{IEEEtran}
\IEEEoverridecommandlockouts

\usepackage{amsmath,amssymb,amsfonts}
\usepackage{algorithmic}
\usepackage{graphicx}
\usepackage{textcomp}
\usepackage{xcolor}
\usepackage[hidelinks]{hyperref}
\usepackage{listings}
\usepackage{xcolor}
\usepackage{tabularx}
\usepackage{booktabs}

\definecolor{backcolour}{rgb}{245,245,245}

\lstdefinestyle{mystyle}{
    backgroundcolor=\color{backcolour},
    numbers=left
}

\usepackage[style=ieee,doi=false,isbn=false,alldates=iso]{biblatex}
\begin{document}

\title{Trusting a Smart Contract Means Trusting Its Owners: Understanding Centralization Risk
\thanks{This work was supported by the Algorand Centres of Excellence programme managed by Algorand Foundation. Any opinions, findings, and conclusions or recommendations expressed in this material are those of the author(s) and do not necessarily reflect the views of Algorand Foundation.}
}

\author{\IEEEauthorblockN{1\textsuperscript{st} Metin Lamby}
\IEEEauthorblockA{
\textit{Technical University of Munich}\\
Munich, Germany \\
metin.lamby@tum.de}
\and
\IEEEauthorblockN{2\textsuperscript{nd} Valentin Zieglmeier}
\IEEEauthorblockA{
\textit{Technical University of Munich}\\
Munich, Germany \\
valentin.zieglmeier@tum.de}
\and
\IEEEauthorblockN{3\textsuperscript{rd} Christian Ziegler}
\IEEEauthorblockA{
\textit{Technical University of Munich}\\
Munich, Germany \\
christian.ziegler@tum.de}
}

\maketitle

\begin{abstract}
Smart contract access control mechanisms can introduce centralization into supposedly decentralized ecosystems.
In our view, such centralization is an overlooked risk of smart contracts that underlies well-known smart contract security incidents.
Critically, mitigating the known vulnerability of missing permission verification by implementing authorization patterns can in turn introduce centralization.
To delineate the issue, we define centralization risk and describe smart contract source code patterns for Ethereum and Algorand that can introduce it to smart contracts.
We explain under which circumstances the centralization can be exploited.
Finally, we discuss implications of centralization risk for different smart contract stakeholders.
\end{abstract}


\section{Introduction}
One core goal of ecosystems leveraging permissionless blockchain technology is decentralization.
However, studies on major blockchain systems indicate that this property is currently not fully achieved~\cite[e.g.,][]{Sai2021TaxonomyReview}.
For example, in Proof of Work (PoW) blockchains, the majority of mining resources are concentrated in a small number of nodes or mining pools~\cite{Beikverdi2015TrendNetwork}. Centralization also occurs in governance, for example in the decision of core developers to lower the minimum transaction fee in Bitcoin~\cite{Gervais2014IsCurrency}.
Similarly, decentralized finance (DeFi), specifically smart contracts, promises to be of use when conducting transactions without having to rely on a central entity within the ecosystem.
Smart contracts can be leveraged as agreements between mutually distrusting participants.
They are automated by the consensus mechanism of the underlying blockchain, removing the role of a trusted authority. 

In this paper, however, we argue that smart contracts themselves can be a source of centralization.
Privileged authorization to contract functionality is implemented with access control patterns. Even though restricting access appears to be a logical step in an open ecosystem~\cite{Chen2021ASecurity},
it may harm its decentralized nature. The dilemma between authorization and centralization sources implemented into a smart contract appears to be a question of priorities.
As this may impose risks on stakeholders, the dilemma needs to be discussed. 

Since DeFi is a developing and open ecosystem, it is challenging to protect all stakeholders.
Therefore, many security incidents occur, including unexpected financial loss to users, liquidity providers, and other parties~\cite{Zhou2022SoK:Attacks}. Stakeholders suffered a total loss of at least US\$ 3.24 billion after interacting with the DeFi ecosystem from Apr 30, 2018 to Apr 30, 2022~\cite{Zhou2022SoK:Attacks}. The most common DeFi incident cause are smart contract vulnerabilities~\cite{Zhou2022SoK:Attacks, Chen2021ASecurity}, including hacks due to user privileges.

\section{Related Work}

The code pattern of introducing privileged parties into smart contracts is coined as a ``tainted owner variable'' vulnerability by~\textcite{Brent2020Ethainter:Vulnerabilities} and an ``overpowered role'' vulnerability in the GitHub repository of~\textcite{Tikhomirov2018SmartCheck} which is part of the publication itself. The issue exists if a function is callable by a single privileged user, resulting in the underlying contract having a dependency on the respective user address. Undesirable consequences for investors may occur if the private key of the superuser address becomes compromised~\cite{SmartCheck2019SmartCheckRoles}. 

To detect such issues in Ethereum ERC token contracts, \textcite{Ma2022Pied-Piper:Contracts} provide an application that leverages a hybrid analysis method integrating datalog analysis and directed fuzzing. Furthermore, \textcite{MichaelFrowisDetectingEthereum} use symbolic execution to reveal ownership structures in any given Ethereum smart contract.

\section{Understanding Centralization Risk}

In the following, we first define centralization risk.
Then, we outline how access controls can create and exacerbate it.
Finally, we consider the relevance for research and practice.
We explain the importance of centralization risk in practice and highlight an identified awareness gap in academia.

\subsection{Definition}
We define centralization-related risks in smart contracts based on academic research and the practitioner's perspective.
To uncover relevant sources, we used results from both a systematic and an exploratory literature review.
In academia, \citeauthor{Ma2022Pied-Piper:Contracts} denote centralization risk as ``backdoor threats'' and define them as ``threats related to high-privileged functions''~\cite[1--2]{Ma2022Pied-Piper:Contracts}.
On the practicioner's side, the smart contract auditor \emph{Certik}
classifies the vulnerability with major severity and defines findings related to centralization as application logic that acts against the nature of decentralization, including ``specialized access roles in combination with a mechanism to relocate funds''~\cite{CertikFidelisReport}. Finally, \emph{Coinbase}
defines the issue as ``superuser risks'' and flags their occurrence when ``single actors have the sole authority to execute a dangerous function''~\cite{TheCoinbaseDigitalAssetProtocolSecurityTeam2022HowRisks}.
We can therefore define centralization risk in smart contracts as the source code including privileged access patterns on fund-modifying logic.

\subsection{Access Controls Becoming Risks}
The ``access control'' vulnerability, caused by inadequate authentication enforced by a contract on critical contract functionality~\cite{Chen2021ASecurity},
conflicts with centralization risk.
Implementing access control mechanisms \emph{introduces} user privileges into smart contracts.
Therefore, they are an important factor that creates sources of centralization.
This can implicate risks, especially when private keys of privileged users are leaked~\cite{Chen2021ASecurity}.
If the private key of a privileged user's address is compromised, an attacker can take control of critical contract functionality, triggering undesirable actions for stakeholders. An example of a potential attack goal is the artificial dilution or inflation of the value of a token implemented as a smart contract~\cite{Brent2020Ethainter:Vulnerabilities}.
Therefore, negligent private key management can make authorization patterns vulnerable even though they might initially be implemented to reduce contract risks.
For example, consider the Ronin hack: In March 2022, Axie Infinity’s Ronin Network experienced a hack that resulted in the theft of US\$ 625 million. The attackers compromised a privileged user’s private keys and conducted restricted withdrawals on the Ronin bridge. This lead to the Axie DAO validator signature to be abused, resulting in the theft of funds. The following network disruption affected both the Ronin Bridge and the Katana automated market marker~\cite{Scharfman2023Decentralized2}.

Because of the interplay between authorization patterns and private keys granting access, we argue that centralization risk are multidimensional, consisting of an implementation and a social component. Subsequently, authorization concepts should be evaluated carefully before contract deployment. Yet, if avoided, other vulnerabilities might emerge, including the risk of making applications self-destructible~\cite{Brent2020Ethainter:Vulnerabilities}. Which risk to accept becomes a question of priorities as we further discuss below.

\subsection{Relevance in Practice}
Major players from the practitioners’ community highlight the importance of centralization risk.
Their caution is warranted, as centralization issues were the most common attack vector exploited in 2021~\cite{CertikThe2021}.
According to the report, US\$ 1.3 billion in user funds were lost across 44 DeFi hacks due to user privileges in 2021.

\subsection{Awareness Gap in Academia}
Even though the practical relevance of centralization risk is clear, we find that there is an awareness gap in academia.
The vulnerability is not mentioned in any academic paper on Algorand smart contract vulnerabilities that we screened in an exploratory review of the field.
When reviewing the literature on Ethereum smart contract vulnerabilities, we only found two papers that mention it, namely~\cite{Ma2022Pied-Piper:Contracts, Brent2020Ethainter:Vulnerabilities}. The lack of academic research on this vulnerability suggests low relevance.
However, as we note above, the risks are of significant importance in practice.
Therefore, we find a delta in the suggested relevance of centralization risk between research and practice.
This suggests a lack of attention in academia.

\section{Detecting Centralization Risk}\label{detectingCentraRisks}

Static analysis can be leveraged to detect patterns in smart contracts that contribute to a potential centralization risk-related vulnerability.
In the following, we show detection patterns for Ethereum (Solidity) and Algorand (TEAL).

\subsection{Solidity Patterns and Slither Detector}
There are different Solidity source code patterns with which \emph{access control} to contract functions can be implemented, as illustrated in table \ref{centralizationRisksSolidity}. Regardless of the pattern used, Solidity access control implementations usually check whether the global variable \texttt{msg.sender}, which allows contract developers to read the address of the sender of the current contract call, satisfies certain conditions~\cite{Xue2020Cross-contractContracts}. In addition to access control patterns, there is one common Solidity source code pattern which is often used in smart contracts to \emph{relocate funds}.

\begin{table}[htpb]
  \caption{Solidity centralization risk related vulnerability patterns}\label{centralizationRisksSolidity}
  \centering
  \begin{tabular}{m{1.5cm} m{6.4cm}}
    \toprule
      \textbf{Pattern} & \textbf{Code Example} \\
    \midrule
        onlyOwner modifier~\cite{FerreiraTorres2020GIS:Attacks, Schiffl2021TowardsControl, Brent2020Ethainter:Vulnerabilities, Ma2022Pied-Piper:Contracts} & 
        
        \begin{lstlisting}[numbers=none]
modifier only_owner {
    require(msg.sender == address(...));
    ...
}

function fun(...) only_owner ... { ... }
        \end{lstlisting}
      \\
      \midrule
      
      require within function~\cite{Xue2020Cross-contractContracts, FerreiraTorres2022Elysium:Contracts, Schiffl2021TowardsControl} & 
        
        \begin{lstlisting}[numbers=none]
function fun() ... {
    require(address(...) == msg.sender)
    ...
}
        \end{lstlisting}
      \\
      \midrule
      
      if statement within function~\cite{Schiffl2021TowardsControl} & 
        \begin{lstlisting}[numbers=none]
function fun() ... {
    if (msg.sender == address(...)) {
        ...
    }
}
        \end{lstlisting}
      \\
      \midrule
      
      balance modification~\cite{Wohrer2018SmartSolidity, Feist2019Slither:Contracts} & 
        \begin{lstlisting}[numbers=none]
mapping(address => uint) bals;

function fun (...) ... {
    bals[...] = bals[...].add(...);
}
        \end{lstlisting}
      \\
      
    \bottomrule
  \end{tabular}
\end{table}

\subsection{TEAL Patterns and Tealer Detector}
In order to determine how to detect centralization in TEAL smart contracts, we take advantage of non-academic sources as well as compilation of PyTEAL patterns. First, the Algorand GitHub\footnote{Algorand GitHub: \url{https://github.com/algorand/smart-contracts}} 
provides TEAL contracts that we can use to identify access control and balance modification patterns in TEAL. In addition, \textcite{BaratellaMatteo2022DecentralizedBlockchain} implements an authentication logic in PyTEAL.
From those, we derive the patterns in \ref{centralizationRisksTEAL}.

\begin{table}[htpb]
  \caption{TEAL centralization risk related vulnerability patterns}\label{centralizationRisksTEAL}
  \centering
  \begin{tabular}{m{3.5cm} m{4.4cm}}
    \toprule
      \textbf{Pattern} & \textbf{Code Example} \\
    \midrule
        address comparison with assert~\cite{AfricaCodeAcademyFidelisProgram, BaratellaMatteo2022DecentralizedBlockchain} & 
        
        \begin{lstlisting}[numbers=none]
byte "manager"
app_global_get
txn Sender
==
assert
        \end{lstlisting}
      \\
      \midrule
      
      address comparison with branch opcodes~\cite{AlgorandAlgorandContract, AlgorandAlgorandContractb} & 
        
        \begin{lstlisting}[numbers=none]
byte "Creator"
app_global_get
txn Sender
==
...
bz failed
        \end{lstlisting}
      \\
      \midrule
      
      balance modification~\cite{AlgorandAlgorandContract} & 
        \begin{lstlisting}[numbers=none]
...
byte "MyBalance"
...
app_local_put
        \end{lstlisting}
      \\
      
    \bottomrule
  \end{tabular}
\end{table}

\section{Implications}\label{implications}
There are multiple implications that arise from this risk, affecting investors, developers, but also CEXs. 

\subsection{Trust Assumptions for Investors}
Even though smart contract code is automatically run by the consensus mechanism of a blockchain without a trusted authority, 
the necessity for trust remains.
DeFi aims to eliminate single points of attack that exist in the traditional financial ecosystem~\cite{Schar2020DecentralizedMarkets}, but they can be reintroduced by access control patterns on critical smart contract functionality. Therefore, while the enforcement of business logic might be distributed, stakeholders could depend on trusted authorities to execute contract functionality in their favor. As a result, investors engaging in tokenized assets may be exposed to trust assumptions~\cite{RobBehnke2022DesigningContracts} that are difficult to evaluate. If exploited, the vulnerability can cause token holders to incur a loss on their investment.
While the reliable parties in CeFi are the intermediaries, privileged access to critical smart contract functionality creates a dependency on a small set of accounts within the DeFi building blocks.
This seems to be a questionable ideal as the notion of decentralization loses validity. 

\subsection{Implementation Dilemma for Developers}\label{dilemma}
Contract developers are exposed to the dilemma of implementing access control patterns and introducing overpowered user roles. Whether to restrict a function becomes a question of priorities affecting many stakeholders. In addition, the concept of multiple signature addresses may gain relevance for contract developers. Multisignature (multisig) wallets refer to a ``type of account that requires a minimum number of addresses to sign a transaction before executing it''~\cite{RobBehnke2022DesigningContracts}. Using a multisig for access control introduces an extra layer of security, since operations on the underlying contract require consent from multiple parties~\cite{Ethereum.org2023SmartSecurity}. This makes it more difficult for a malicious user to manipulate sensitive contract functions, whether the user is the owner of a contract or is assigned a specific role. Hence, decentralization is increased, as no single entity has total control of a contract function~\cite{RobBehnke2022DesigningContracts}. This technique divides responsibility for key management between multiple parties and prevents the loss of a single private key from potentially causing undesirable and irreversible consequences~\cite{Ethereum.org2023SmartSecurity}. Nevertheless, the multisig concept reduces centralization risk but does not mitigate the problem as private key leakages of multiple privileged users have caused smart contract exploits in the past. Also, the mentioned benefits come with the cost of worse usability and increased reaction time when signing a transaction. 

\subsection{Increased Security Challenges for CEXs}\label{manualEffort}
Further affected stakeholders of this discussion might be CEXs that enable the trade of tokens, their affiliated custody providers, and smart contract auditors. CEXs implement listing processes that aim to identify vulnerable token contracts to prevent them from being listed. Usually, exchanges hire third-party token audits to provide that service. Often, auditors leverage automated smart contract vulnerability detection tools to increase workflow efficiency. Nevertheless, the social component of centralization risk makes it impossible for auditors to detect the vulnerability automatically.
Detection tools can only determine the existence of implementation patterns that \emph{may cause} the vulnerability. However, since contracts containing superuser roles on critical functions are said to be as secure as the protections on those roles~\cite{TheCoinbaseDigitalAssetProtocolSecurityTeam2022HowRisks}, they need to be evaluated manually.
Therefore, smart contracts may not only need to be audited on a source code level but also on a social level, represented by the roles that have access to restricted function execution. A know your customer (KYC) process including an audit on private key management might be an initiative to evaluate the social risk vector a smart contract possesses. As a result, CEXs might introduce additional burdens for token listings while prioritizing those parties admitting their tokens for listing that have a good reputation, possibly bigger institutions. If the addressed risk is neglected, possible consequences might include regulatory scrutiny and reputational damage.   

\section{Conclusion}
Smart contract access control mechanisms can introduce sources of centralization into supposedly decentralized ecosystems, including DeFi.
We observe a delta in the perceived relevance of centralization risk between academia and industry, suggesting a research gap.
We address this gap by
(1) identifying centralization risk and its characteristics and
(2) presenting patterns to detect it with static analysis.
Our findings implicate increased trust assumptions for token investors, greater development efforts and responsibility for software engineers, and security challenges for CEXs. 
However, centralization risk is multidimensional and includes a social component.
While source code patterns that enable the vulnerability can be detected automatically, the private key management of authorized users requires manual security audits. 

To conclude, the introduction of privileged access to functionality within a smart contract can create single points of attack, leading to undesirable consequences if exploited.
Therefore, we argue that trusting a smart contract means trusting its ``owners,'' meaning the privileged stakeholders as well as their key management.

\printbibliography{}

\end{document}